\documentclass[twocolumn,prb]{revtex4}

\usepackage{dcolumn}
\usepackage[xdvi]{graphicx}
\usepackage{amsmath}
\usepackage{amssym}

\begin{document}
\title {Correlation versus mean-field contributions to excitons,
multi-excitons, and charging energies in semiconductor quantum dots}
\author{J.~Shumway, A.~Franceschetti, and Alex Zunger}
\affiliation{National Renewable Energy Laboratory, Golden, Colorado 80401}
\date{October 3, 2000}
\begin{abstract}
Single-dot spectroscopy is now able to resolve the energies of 
excitons, multi-excitons, and charging of semiconductor quantum dots with
$\alt 1$~meV resolution.  We discuss the physical content of these energies
and show how they can be calculated via Quantum Monte Carlo (QMC)
and Configuration Interaction (CI) methods.  The spectroscopic energies
have three pieces: (i) a ``perturbative part'' reflecting
carrier-carrier direct and exchange Coulomb energies obtained from
{\em fixed} single-particle orbitals, (ii) a ``self-consistency correction'' 
when the single particle orbitals are allowed to adjust to the
presence of carrier-carrier interaction, and (iii) a ``correlation correction.''
We first apply the QMC and CI methods to a modle single-particle Hamiltonian:
a spherical dot with a finite barrier and {\em single-band} effective mass.
This allows us to test the convergence of the CI and to establish the 
relative importance of the three terms (i) -- (iii) above.  Next, we
apply the CI method to a realistic single-particle Hamiltonian
for a CdSe dot, including via a pseudopotential description the atomistic
features, {\em multi-band coupling}, spin-orbit effects, and surface 
passivation.  We include all bound states (up to 40,000 Slater determinants)
in the CI expansion.
Our study shows that: (1) typical exciton transition energies, which
are $\sim 1$~eV, can be calculated to better than 95\% by
perturbation theory, with only a $\sim 2$~meV correlation correction;
(2) typical electron addition energies are $\sim 40$~meV, of 
which correlation contributes very little ($\sim 1$~meV); 
(3) typical biexciton binding energies are positive and $\sim 10$~meV
and almost entirely due to correlation energy,
and exciton addition energies
are $\sim 30$~meV with nearly all contribution due to correlation;
(4) while QMC is currently limited to a single-band effective
mass Hamiltonian, CI may be used with much more realistic models,
which capture the correct symmetries and electronic structure of the
dots, leading to qualitatively different predictions than effective mass
models; and (5) and CI gives excited state energies necessary to identify
some of the peaks that appear in single-dot photoluminescence spectra.
\end{abstract}
\maketitle

\section{Introduction: The physical content of exciton, multiexciton,
and charging energies in dots}

Small semiconductor dots, such as semiconductor embedded Stranski-Krastanow
(SK) dots or ``free-standing'' colloidal dots, are engineered and studied for
their optical and transport properties.\cite{Bimberg:1999,Woggon:1997,%
Jacak:1998} Measurements on these dots have centered around quantities 
such as excitons,\cite{Marzin:1994,Fafard:1995,Yang:2000}
multi-excitons,\cite{Dekel:1998,Dekel:2000,Landin:1999,Toda:1999,Zrenner:2000}
and charging energies.\cite{Fricke:1996,Tarucha:1996,Banin:1999,%
McEuen:1991,McEuen:1992}
Advanced experimental techniques, such as 
single-dot spectroscopy, are able to resolve 
to such energies to $\alt 1$~meV resolution.  
This article discusses the physical content of such measured quantities 
in terms of the mean-field (direct and exchange) Coulomb  energies,
which are relatively simple to model, and correlation energies, which
we calculate by two leading methods in the field --- Quantum Monte Carlo
(QMC) and Configuration-Interaction (CI).

Let us consider a quantum dot with $M$ holes in the valence band and
$N$ electrons in the conduction band.  The total energy of the dot
is $E_{M,N}(\alpha)$, where $\alpha$ is a quantum number
that identifies the state of the system.  Only differences in energy
are accessible to experiment.  We focus on four physical quantities:

\paragraph*{(i) Exciton energies.}
The {\em exciton transition energy} $E_X^{(ij)}$
is the difference in total energy of a dot having as a dominant
configuration an electron in level $e_i$ and a hole in level $h_j$ and
a dot in the ground state,
\begin{equation}
\label{eq:extran}
E_X  = E_{1,1}(e_i^1,h_j^1) - E_{0,0}.
\end{equation}
Typical excitonic transition energies
in III-V or II-VI dots, measured experimentally\cite{Fafard:1995,Bawendi:1990}
by photoluminescence (PL) or by absorption, are 1 -- 3 eV.
The {\em exciton binding energy} $\Delta_X$ is the difference 
between the total energy of 
a system consisting of two infinitely separated identical dots,
one with a hole in $h_0$ and the other with an electron in $e_0$,
and the total energy of a quantum dot with an exciton:
\begin {equation}
\label{eq:exciton}
\Delta_X = E_{1,0} + E_{0,1} - E_{1,1} - E_{0,0},
\end {equation}
where $E_{1,0}$ stands for $E_{1,0}(h^1_0e^0_0)$,
$E_{0,1} = E_{0,1}(h^0_0e^1_0)$, and $E_{1,1} = E_{1,1}(h^1_0e^1_0)$.
Typical exciton binding energies in III-V and II-VI dots
are 10-200~meV.\cite{Fafard:1995,Bawendi:1990}

\paragraph*{(ii) Biexciton energies.}
The {\em biexciton binding energy} $\Delta_{XX}$
is the difference between 
twice the exciton energy (or the energy
of a system of two infinitely separated
dots, each with an electron-hole pair),
and the biexciton energy:
\begin{equation}
\label{eq:biex}
\Delta_{XX} = 2E_{1,1} - E_{2,2} - E_{0,0}.
\end{equation}
The biexciton binding energy is positive 
(``bound biexciton'') when the total energy of two
excitons in the same dot is lower than the energy of the two
excitons in two separate dots.  A bound biexciton appears as a red-shifting of
the exciton luminescence energy when a second exciton is present.
This was seen in single-dot spectroscopy e.~g.~for InAs/GaAs.\cite{%
Dekel:1998,Dekel:2000,Zrenner:2000,Zrenner:1999}
Biexciton binding energies in III-V dots are 
1 -- 6~meV.\cite{Zrenner:1999,Brunner:1994,Kuther:1998,Kamada:1998,%
Kulakovskii:1999,Gindele:1999}

\paragraph*{(iii) Multi-exciton energies.}
The {\em N-th exciton charging energy $W_{N}$} is
the minimum energy needed to add to a dot having $N-1$ electron-hole pairs 
(excitons) in their ground state one additional exciton,
\begin{equation}
\label{eq:ncharge}
W_{N} = E_{N,N} - E_{N-1,N-1}.
\end{equation}
Physically, $W_N$ is the highest possible energy for a photon emitted
in the transition from the lowest energy state of $N$ excitons
to a state with $N-1$ excitons.  The difference in successive
multi-exciton charging energies is the {\em N-th exciton addition
energy} $\Delta^{(X)}_{N,N+1}$,
\begin{equation}
\begin{split}
\label{eq:nadd}
\Delta_{N,N+1}^{(X)} & = W_N^{(X)} - W_{N-1}^{(X)}  \\
& = E_{N+1,N+1} +E_{N-1,N-1} - 2E_{N,N}.\\
\end{split}
\end{equation}

\paragraph*{(iv) Electron loading energies.}
The {\em electron charging energy} $\mu_{N}^{(e)}$ is the 
chemical potential needed to add an electron to 
a dot already having $N-1$ electrons:
\begin{equation}
\label{eq:qcharge}
\mu_{N}^{(e)} = E_{0,N} - E_{0,N-1},
\end{equation}
whereas the {\em electron addition energy} is the difference between two
successive chemical potentials,
\begin{equation}
\begin{split}
\label{eq:qadd}
\Delta_{N,N+1}^{(e)} & = \mu_N^{(e)} - \mu_{N-1}^{(e)} \\
                     & = E_{0,N+1} +E_{0,N-1} - 2E_{0,N}. \\
\end{split}
\end{equation}
Electron addition energies in colloidal dots\cite{Banin:1999} are 
$\sim 200$~meV, whereas in SK dots\cite{Fricke:1996,Warburton:1998}
they are $\sim 20$~meV.

\begin{figure}
\includegraphics[width=\linewidth]{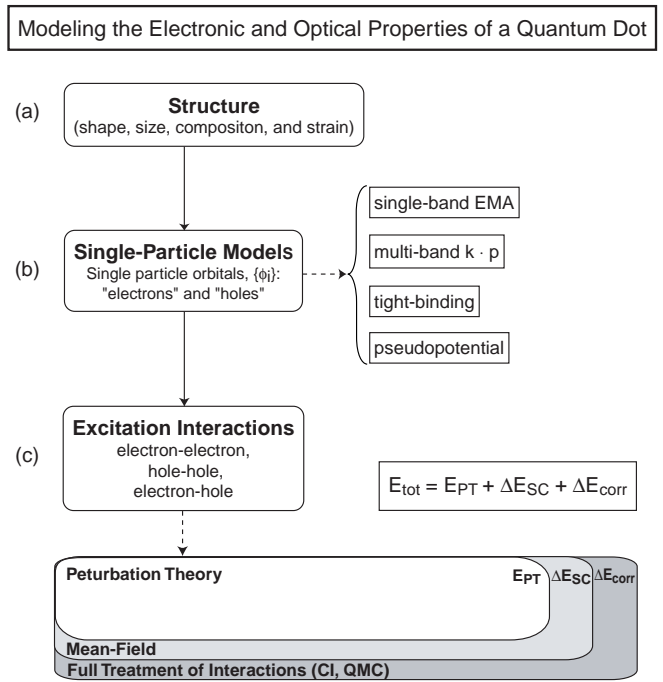}
\caption{The three steps to modeling a quantum dot.
(a) The structure is modeled by choosing a size, shape, and composition
profile, and determining the strain.
(b) Single particle properties of the electrons and holes are found
by solving a Schr\"odinger equation for a chosen level of
renormalization (EMA, $k\cdot p$, tight-binding, or pseudopotential).
(c) Interactions between excitations (electron-electron, electron-hole,
hole-hole) are added to the single particle model, using either
perturbation theory $E_{\text{PT}}$; self-consistent mean-field theory, 
which adds the self-consistent contribution $\Delta E_{\text{SC}}$, or full
treatment (CI or QMC), which adds the correlation energy $E_{\text{corr}}$.}
\label{fig:methods}
\end{figure}

The definitions given here in Eqs.~(\ref{eq:extran})--(\ref{eq:qadd}) are
operational, model-independent.  A central question in the field
is how to approximate these quantities through models.  This requires knowing
how much of the energy involved in the precesses described by
Eqs.~(\ref{eq:extran})--(\ref{eq:qadd}) are due to ``mean-field'' effects,
which can be modeled relatively simply, and how much is due to inter-particle
correlation, which is more intricate to model.   Figure~\ref{fig:methods}
illustrates the steps required to model the 
electronic and optical properties of a quantum
dot: (a) choosing a structure (including size, shape, composition,
and strain), (b) solving a single particle model, and
(c) treating interactions among the electrons and holes.  In this paper
we are concerned with {\em general} trends in correlation in dots, so
we focus mainly on the choice of single particle model, 
Fig.~\ref{fig:methods}(b), and treatment of interactions,
Fig.~\ref{fig:methods}(c).

As illustrated in Fig.~\ref{fig:methods}(b),
the calculations of the quantities of Eqs.~(\ref{eq:extran})--(\ref{eq:qadd})
require one to assume an underlying single-particle model, which determines
the single particle states (conduction electrons and holes).
The single-particle model is cast as a Schr\"odinger equation with an
effective single-particle potential.  This potential
contain all structural information
about the system: the size, shape, composition, surfaces, interfaces of
the dot system.  Various levels of renormalization exist for the quantum 
dot single-particle model.  The simplest is an effective mass
(``particle-in-a-box'') model, in which the electron and hole excitations
come from single parabolic bands.  Better approximations are
the multi-band $k\cdot p$, tight-binding, and pseudopotentials.

The single-particle models do not usually contain the Coulomb interactions
between the single-particle excitations (i.{} e.{} electron-electron, 
electron-hole, and hole-hole excitations).  Instead, these interactions must be
added to the model, as shown in Fig.~\ref{fig:methods}(c).  
We classify the treatment of interaction 
among the single-particle excitations in three levels:
(i) {\em first order Perturbation Theory (PT)},\cite{Dekel:1998,Barenco:1995,%
Franceschetti:1999,Franceschetti:2000a,Franceschetti:2000b,Williamson:2000c}
which includes direct and exchange Coulomb interactions,  $J$ and $K$, 
evaluated from {\em fixed} single-particle orbitals;
(ii) {\em Self-Consistent Mean-Field (MF)},\cite{Franceschetti:1997} in 
which the direct and exchange Coulomb terms are solved self-consistently
[the difference between (ii) and (i) is called ``self-consistency
correction'' $\Delta E^{\text{SC}}$]; and 
(iii) {\em correlated methods, such as CI\cite{Dekel:1998,Barenco:1995,%
Franceschetti:1999,Franceschetti:2000a,Franceschetti:2000b,Williamson:2000c}
or QMC},\cite{Harju:1999,Austin:1988,Bolton:1996,Shumway:1999,Pederiva:2000,%
Lee:1998a,Luczak:2000,Mak:1998,Egger:1999a,Egger:1999b,Egger:1999c} 
which include all many-body effects of interactions.  
The difference between the exact energy (iii) and the mean-field energy (ii)
is called the ``correlation correction,'' $\Delta E^{\text{corr}}$.  Thus, 
the energy for a dot with $M$ holes and $N$ electrons can be separated 
into three terms,
\begin {equation}
\label{eq:threeterms}
E_{M,N}^{\text{tot}} = E_{M,N}^{\text{PT}} + \Delta E_{M,N}^{\text{SC}} 
              + \Delta E_{M,N}^{\text{corr}}
\end {equation}
which are perturbation theory $E_{M,N}^{\text{PT}}$,
self-consistent corrections $\Delta E_{M,N}^{\text{PT}}$, and the
correlation correction $\Delta E_{M,N}^{\text{corr}}$.

\begin{table}
\squeezetable
\caption{Relationship between the choice of single particle
models for quantum dots and the availability of QMC and CI methods to calculate 
correlation energy [see Fig.~\ref{fig:methods}, (b) and (c)].
This information motivates our approach to studying correlation:
first we test the convergence of CI against QMC calculations using a 
simple single-band EMA model, then we present CI calculations on a 
realistic multi-band pseudopotential quantum dot model to illustrate features
missed by the simple model.}
\label{table:civsqmc}
\begin{ruledtabular}
\begin{tabular}{l l c c}
Level of Renormalization &
Model & CI\footnote{While CI may be applied to any model,
it is often under converged.}
 & QMC\\
\colrule
All electron         & Exact Hamiltonian          & no\footnote{Possible
for very small clusters of less than 100 atoms ($R<10$~\AA).} & no\\
\\
Valence only         & Multi-Band Pseudopotential & yes & no$^{\text{b}}$\\
                     & Tight-binding              & yes & no$^{\text{b}}$\\
\\
Active electron only & Multi-Band $k\cdot p$      & yes & no\\
                     & Single-Band EMA            & yes & yes\\
\end{tabular}
\end{ruledtabular}
\end{table}

Due to computational limitations, the methods available to 
calculate correlation are dependent on which single-particle model
is chosen (level of renormalization).
The computational cost for accurately calculating correlation energies
increases rapidly with the number of electrons one needs to consider.
The number of electrons depends on both the dot's size  and
on the type of renormalization one uses for the Hamiltonian.
As summarized in Table~\ref{table:civsqmc}, three levels of renormalization 
are pertinent:

(a) The all electron approach, where the number of electrons
per atom equals its atomic number.  Thus, Si has 14 electrons per atom,
and a 40~\AA{} diameter spherical Si dot has $1,600\times 14 = 22,400$~
electrons.  This is outside the reach of QMC, CI, and density functional
methods.

(b) The valence-only pseudopotential approach, where the ``core''
electrons are removed as dynamic variables and replaced by an
(often non-local) ionic potential.  Thus, Si has 4 electrons per atom,
and a 40~\AA{} diameter spherical Si dot has $1,600\times 4 = 6,400$~
electrons.  This is outside the reach of density functional methods, and
too large for QMC calculations, which are 
currently limited to about 25 Si atoms (100 electrons).\cite{Mitas:2000}
Note that the all-valence pseudopotential approach can be
further simplified, with no additional approximations by searching 
for eigensolutions in a fixed ``energy window,''\cite{Wang:1994a,%
Wang:1994c,Wang:1996a} e. g., near the band edges.
Thus, a 40~\AA{} diameter Si dot would require calculating
$\sim 10$ eigensolutions.  This trick makes pseudopotential calculations
of dots feasible,\cite{Franceschetti:1999,Franceschetti:1997,%
Fu:1998a,Wang:1998a,Wang:2000a} and CI may be used to compute
correlation energies from the single particle solutions.\cite{%
Franceschetti:1999,Franceschetti:2000a,Franceschetti:2000b,Williamson:2000c}
It would be interesting if such folding techniques could be applied
to QMC.

(c) The ``active-electron-only'' Effective Mass Approximation (EMA) approach,
where all of the ``indigenous'' core and valence electrons are eliminated
(replaced by dielectric screening) and only {\em additional}, band-edge 
electrons and holes are considered.  Thus a 40~\AA{} diameter Si dot has
zero electrons.  One can study {\em added} electrons and holes.
This renormalization represents a severe approximation with respect
to levels (a) and (b) above.  Both QMC and CI methods may be readily
applied to EMA Hamiltonians.  Some improvement can be made
by using several bands to describe the additional electrons and holes using
the $k\cdot p$ formalism,\cite{Stier:1999,Grundmann:1995}
but current QMC methods do not treat $k\cdot p$ Hamiltonians.

Most correlated calculations on quantum dots 
have used such a single-band effective mass model [level (c), above],
where multi-band and inter-valley couplings are ignored.
This particle-in-a-box description of the mean-field problem was 
recently\cite{Franceschetti:1999,Franceschetti:1997,Fu:1998a,Wang:1998a,%
Wang:2000a} contrasted with the pseudopotential solution of the 
problem [(b) above] both for ``free-standing'' (colloidal)
dots and for semiconductor-embedded SK dots.
It was found that for ``free-standing'' dots (InP,\cite{Fu:1998a}
CdSe\cite{Wang:1998a}) the effective mass approach can lead to energy 
shifts of the order\cite{Fu:1998a,Wang:1998a} $\sim500$~meV;
lead to reverse order of (s,p) levels;\cite{Fu:1998a}
miss more than half of the single-particle eigenvalues in a 0.5~eV 
energy range near the band edge;\cite{Wang:1998a}
underestimate the Coulomb integrals $J_{ij}$ [Eq.~(\ref{eq:jk})]
by\cite{Franceschetti:1997} $\sim20\%$; and miss 
all the long-range part of the exchange integrals\cite{Franceschetti:1999}
$K_{ij}$.  For pyramidal SK dots\cite{Wang:2000a}
the errors are somewhat smaller: shifts in
the energy levels for electrons and holes are $\sim 35$~meV and $\sim 110$~meV,
respectively; energy spacings from EMA are about a factor of two too large;
and the polarization ratio for dipole transitions along the 
two directions is 1 instead of 1.3.
Such limitation in the EMA create a dilemma when modeling correlation
as summarized in Table~\ref{table:civsqmc}.
On one hand  CI expansions maybe applied
to realistic single-particle models (e.~g. pseudopotentials),
but converge slowly with the number of configurations.
On the other hand, QMC methods can give 
numerically exact answers including all correlation,
but currently are limited to simple single-band effective mass models.
This situation prompts us to use the following strategy to study correlation 
effects: First, we consider a simplified ``particle-in-a-box'' single-band EMA
model  which can be treated both via QMC and CI.  Our best CI calculations
for the EMA model include all bound states, but neglect continuum states.
Second, we consider a CdSe dot whose single-particle 
properties are described realistically by pseudopotentials, and the
correlation is treaded via CI only.  

\begin{table*}
\caption{Measurable quantities for our single-band 
spherical model dot,
with effective masses $m_e=0.1$ and  $m_h=0.5$,
dielectric constant $\epsilon=12$,  
dot-material band gap $E_{\text{gap}}= 1$~eV,
and band offsets $\Delta E_v = 200$~eV
and $\Delta E_c = 400$~meV.
For each quantity we give the magnitude, the mean-field
value, the correlation correction, and the percent of the energy
recovered by CI expansion of using all bound states.
All energies are given in meV, and electron charging and total energies
are measured relative to the dot material CBM.}
\label{table:quantities}
\begin{ruledtabular}
\begin{tabular}{ldddd}
\multicolumn{1}{l}{Quantity} &
\multicolumn{1}{c}{Magnitude} &
\multicolumn{1}{c}{Mean Field} &
\multicolumn{1}{c}{Correlation} &
\multicolumn{1}{c}{\% CI}\\
\colrule
Exciton {\bf total} energy, $E_{1,1}$ $(e_0^1,h_0^1) $
&1136.3 & 1138.3 & 2.0 & 100.1\\
Biexciton {\bf total} energy, $E_{2,2}$ $(e_0^2h_0^2)$ 
&2266.5 & 2277.3 & 10.9 & 100.2\\
{\bf Total} energy of two electrons, $E_{0,2}$ $(e_0^2)$ 
&335.0 & 335.8 & 0.8 & 100.1\\
\\
Exciton {\bf transition} energy, $E_X$, [Eq.~(\ref{eq:extran})]
&1136.3 & 1138.3 & 2.0 & 100.1\\
Exciton {\bf binding} energy, $\Delta_X$, [Eq.~(\ref{eq:exciton})]
&46.2 & 44.1 & 2.0 & 97.8\\
Biexciton {\bf binding} energy, $\Delta_{XX}$, [Eq. (\ref{eq:biex})]
&6.2 & -0.6& 6.8 & 64.5\\
\\
1st exciton {\bf charging} energy, $W_1^{(X)}$, [Eq. (\ref{eq:ncharge})]
&1136.3 & 1138.3 & 2.0 & 100.1\\
2nd exciton {\bf charging} energy, $W_2^{(X)}$, [Eq. (\ref{eq:ncharge})]
&1130.1 & 1139.0 & 8.9 & 100.2\\
1st exciton {\bf addition} energy, $\Delta^{(X)}_{1,2}$, [Eq. (\ref{eq:nadd})]
&-6.2 & 0.6& 6.8 & 64.5\\
\\
1st electron {\bf charging} energy, $\mu^{(e)}_1$, [Eq. (\ref{eq:qcharge})]
&147.5 & 147.5 & 0.0 & 100.0 \\
2nd electron {\bf charging} energy, $\mu^{(e)}_2$, [Eq. (\ref{eq:qcharge})]
&187.5 & 188.3 & 0.8 & 100.1 \\
1st electron {\bf addition} energy, $\Delta^{(e)}_{1,2}$, [Eq. (\ref{eq:qadd})]
&40.0 & 40.8 & 0.8 & 101.4\\
\end{tabular}
\end{ruledtabular}
\end{table*}

Our single-band EMA dot has been chosen to be representative of 
SK and colloidal dots.  We summarize the properties of our model dot
in Table~\ref{table:quantities}.
We find that for the single-band model dot:

(i) Typical exciton transition energies for our model dots are
$\sim 1$~eV, and typical exciton binding energies are $\sim 50$~meV.
Of this, MF gives $> 95$\% of the binding energy.  Correlation is only 
$\sim 2$~meV, of which QMC provides an accurate solution
Although CI misses half the correlation energy, i.e. $\sim 1$~meV, it still
captures $\sim 98$\% of the total binding energy.

(ii) Typical biexciton transition energies for our 
dots are $\sim 2$~eV and typical biexciton binding energies are $\sim 6$~meV.
The biexciton binding energy from mean-field theory is slightly
negative (unbound biexcitons), so the positive biexciton binding
is in fact due to $\sim 6$~meV of correlation energy.  QMC captures
all the correlation energy, whereas our CI captures only half (about 4 meV),
so that the CI estimate of biexciton binding is only about 65\%
of the true value.

(iii) Typical electron charing energies for our dots are 
$\mu_1^{(e)} \approx 150$~meV, 
relative to the dot material CBM, while addition energies are
$\Delta_{1,2}^{(e)} \approx 40$~meV.
Of this, correlation energy is very small ($\sim 1$~meV), so mean-field 
or even perturbation theory describes dot charging and
addition energies very well.

For our realistic CdSe dot we find that CI can be effectively
combined with accurate pseudopotential description of the MF
problem, thus incorporating surface effects, hybridization, multi-band
coupling.  Furthermore, CI can calculate excited states
easily, thus obtaining the many transitions seen experimentally, rather than 
only ground-state--to--ground-state decay calculated by conventional
QMC (note, however, that extensions of QMC to several excited states
are possible\cite{Ceperley:1988,Bernu:1990}).

\section{Methods of Calculation}

\subsection{Uncorrelated methods: perturbation theory 
and mean field methods}
\label{sec:uncormethods}

The first-order perturbation energy $E_{M,N}^{PT}$
[Eq.~(\ref{eq:threeterms})] can be written analytically as:
\begin{widetext}
\begin {equation}
E_{M,N}^{PT} = E_{0,0} + (\sum_c \varepsilon_c - \sum_v \varepsilon_v )  +
\sum _{v < v'} (J_{v,v'} - K_{v,v'}) +
\sum _{c < c'} (J_{c,c'} - K_{c,c'}) -
\sum _{v,c} (J_{v,c} - K_{v,c}),
\end {equation}
\end{widetext}
where $\varepsilon_i$ are the single-particle energies, $J_{i,j}$ 
are the direct Coulomb
energies, and $K_{i,j}$ are the exchange energies. 
The single-particle energies $\varepsilon_i$ 
are often obtained from the solution of an effective single-particle 
Schr\"odinger equation,
\begin{equation}
\label{eq:se}
\{ -\frac{1}{2}\nabla^2 + V_{\text{eff}} \} 
\psi_i = \varepsilon_i \, \psi_i
\end{equation}
where $V_{\text{eff}}$ is an effective potential.
The Coulomb and exchange energies are given in terms of the single-particle
wave functions $\psi_i$ by:
\begin {eqnarray}
\label{eq:jk}
J_{i,j} &=& \int {|\psi_i ({\bf r})|^2 \, |\psi_j ({\bf r}')|^2 \over 
\epsilon ({\bf r}, {\bf r}') \, |{\bf r} - {\bf r}'|} \, d{\bf r} \, d{\bf r}' 
\nonumber\\
K_{i,j} &=& \int {\psi_i^* ({\bf r}) \, \psi_j^* ({\bf r}) \, \psi_i ({\bf r}') \, \psi_j ({\bf r}')  \over 
\epsilon ({\bf r}, {\bf r}') \, |{\bf r} - {\bf r}'|} \, d{\bf r} \, d{\bf r}',
\end {eqnarray}
where $\epsilon$ is the dielectric constant of the quantum dot. 

The self-consistent contribution $E^{\text{SC}}_{M,N}$, given by the
first two terms on the right hand side of Eq.~(\ref{eq:threeterms}), arises
from the self-consistent rearrangement of the single-particle wavefunction
in respond to the electrostatic field, Eq.~(\ref{eq:jk}),
generated by the excitation of electrons and holes.

\subsection{The correlated, many-particle methods}
\label{sec:cormethods}

\subsubsection{Quantum Monte Carlo}

The original QMC method\cite{McMillian:1965}  was based on the 
variational technique,
a simple, yet powerful theoretical tool.  In a variational calculation,
one proposes a parameterized trial wavefunction $\Psi_T^{\{\lambda\}}(R)$,
where $\lambda$ represents a set of variational parameters and $R$
represents the coordinates of all the particles.
The energy expectation value
\begin{equation}
E^{\{\lambda\}}_T = \frac{\int dR \Psi^{\{\lambda\}*}_T(R) 
                                H \Psi^{\{\lambda\}}_T(R)}
                    {\int dR \Psi^{\{\lambda\}*}_T(R) \Psi^{\{\lambda\}}_T(R)},
\end{equation}
may be minimized with respect to the variational parameters $\lambda$
to give an estimate for the ground state energy and 
ground state wavefunction.
This integral may be evaluated analytically, or Monte Carlo integration
may be used.  In this simplest formulation, QMC is formally equivalent
to the variational techniques commonly applied to excitons in
nanostructures.\cite{Bastard:1988}  Because the integral is over all
electron and hole coordinates $R$, variational QMC calculations resemble
classical simulations: a configuration of particle positions $R$ undergoes
a random walk through configuration space, using the rules of Metropolis
Monte Carlo integration.  The sequence of configurations,
$R_i, R_{i+1}, \ldots$, samples the density $|\Psi_T(R)|^2$.

The real power of QMC is that it can go beyond the variational
formalism and actually project the true ground state energy
from an input variational trial function, $\Psi_T$.\cite{Ceperley:1979}
By weighting
the the configuration as it samples configuration space, 
the random walk can identified
with the imaginary time propagator $\exp(-H\tau)$.  In this diffusion
Monte Carlo algorithm,\cite{Ceperley:1979,Schmidt:1984} the random walk 
in configuration space actually
samples $\Psi_T^* \Phi_0$ where $\Phi_0$ is the true
ground state wavefunction. The energy expectation value along the walk
$E_0 = \langle\Psi_T |H | \Phi_0 \rangle /  \langle\Psi_T | \Phi_0 \rangle$
is then the true ground state energy of the many-body Hamiltonian.
That is, even though the true ground state wavefunction $\Phi_0$ is
never explicitly calculated, its energy can be sampled from a
random walk.  In the remainder of the paper, the term QMC will refer
to the diffusion Monte Carlo algorithm, unless explicitly noted
otherwise.

Applications of QMC to quantum dots have used 
variational QMC,\cite{Harju:1999}
diffusion QMC,\cite{Austin:1988,Bolton:1996,Shumway:1999,Pederiva:2000,%
Lee:1998a,Luczak:2000}
and a path-integral formulation, related to the diffusion algorithm
and based on Feynman path integrals.\cite{Mak:1998,Egger:1999a,Egger:1999b,%
Egger:1999c}
Harju et al\cite{Harju:1999} have used both direct diagonalization
and VMC to calculate the ground state energy of up to 6 electrons in a
two-dimensional harmonically confined dot.
Diffusion QMC within the EMA has been used
(1) by Austin\cite{Austin:1988} to calculate 
the binding energy of excitons in a spherical dot as a function of dot radius,
(2) by Bolton\cite{Bolton:1996} to calculate
the energy of up to 4 electrons in a two-dimensional harmonically confined
dot in the presence of a magnetic field,
(3) by Shumway et al\cite{Shumway:1999} 
to calculate total energies for electron addition to a pyramidal dot,
(4) by Pederiva et al\cite{Pederiva:2000}
to calculate ground and excitation energies for up to 13 electrons 
in a three-dimensional harmonically confined dot and compare to HF and LSDA,
and (5) by Luczak et al\cite{Luczak:2000} to study energies of
up to 20 electrons confined to a two-dimensional harmonic potential.
Lee et al\cite{Lee:1998a} have used QMC within the EMA
to study a pair of electrons in a two-dimensional parabolic confining
potential.
Path integral QMC has been used by Egger et al\cite{Egger:1999a}
to studied crossover from Fermi liquid
to Wigner molecule behavior using PIMC within the EMA on up to 8 electrons in
a two-dimensional harmonically confined dot, and by
Harting et al\cite{Harting:2000} 
to calculate the total energy of up to 12 electrons in a two-dimensional
harmonically confined dot.

\subsubsection{Configuration Interaction}

In the CI approach, the solutions of the many-body Hamiltonian are
expanded in terms of Slater determinants $|\Phi\rangle$
obtained by removing $M$ electrons from the valence band and
adding $N$ electrons to the conduction band:
\begin{equation}
\label{eq:ci}
|\Psi\rangle = \!\!\sum_{h_1\ldots h_M}\!
\sum_{\phantom{h}e_1\ldots e_N}\!\! A(h_1\ldots h_M,e_1\ldots e_N) 
|\Phi_{h_1\ldots h_M,e_1\ldots e_N} \rangle,
\end{equation}
where:
\begin{equation}
\label{eq:configuration}
|\Phi_{h_1\ldots h_M,e_1\ldots e_N}\rangle =
d^\dag_{h_1}\cdots d^\dag_{h_M} c^\dag_{e_1}\cdots c^\dag_{e_N}
\, |\Phi_0 \rangle.
\end{equation}
Here 
$d^\dag_{h_1}\cdots d^\dag_{h_M}$ create {\em holes} in the
valence-band states $h_1\ldots h_M$, 
while $c^\dag_{e_1}\cdots c^\dag_{e_N}$
create {\em electrons} in the conduction band states $e_1\ldots e_N$.
The Hamiltonian is then diagonalized in the basis
of Slater determinants $|\Phi\rangle$.  This approach gives access to
not only the ground state of the system, but also excited states.

Full CI (FCI) includes all possible determinants from a given (finite) set of
single particle basis functions, i. e. $N_h$ hole orbitals and $N_e$ 
electron orbitals.  In the limit of an infinite
set of basis functions, $(N_h,N_e)\rightarrow (\infty,\infty)$,
FCI provides the exact many-body solution, which 
is equivalent to the QMC results.  However, most CI applications use
a small and finite basis set to solve the Schr\"odinger problem.
Thus, even including in the CI expansion
all possible Slater determinants from a finite number of 
single-particle states (FCI) does not provide an exact solution, in contrast
to QMC.  For our calculations we also only use a small, finite basis 
set of bound states, denoted $(N_h,N_e)$, therefore ground state 
total energies from FCI will be 
above the true ground state total energy.
A useful truncated CI basis is Singles and Doubles Configuration Interaction
(SDCI), which the set of all determinants obtained by exciting at most
two particles (electrons or holes) from the ground-state 
(or reference) determinant.  
SDCI is equivalent to FCI for a single exciton (or two electrons), but is 
an approximation for two or more excitons (or three or more electrons).

The CI method has been used in the past to solve the the many-body
Schr\"odinger equation in the EMA approximation\cite{Landin:1999,%
Barenco:1995,Hawrylak:1999,Hawrylak:2000a,Hawrylak:2000b,Brasken:2000} and
also tight binding.\cite{Dib:1999}
More recently, the CI approach has been used in the context
of the empirical pseudopotential method (EPM) for single 
excitons,\cite{Franceschetti:1999} electron and hole addition
energies,\cite{Franceschetti:2000a,Franceschetti:2000b} and
multiexcitons.\cite{Williamson:2000c}

\section{Application of QMC and CI to a single-band effective-mass dot with 
finite barrier}

\begin{figure}
\centerline{\includegraphics[width=0.79\linewidth]{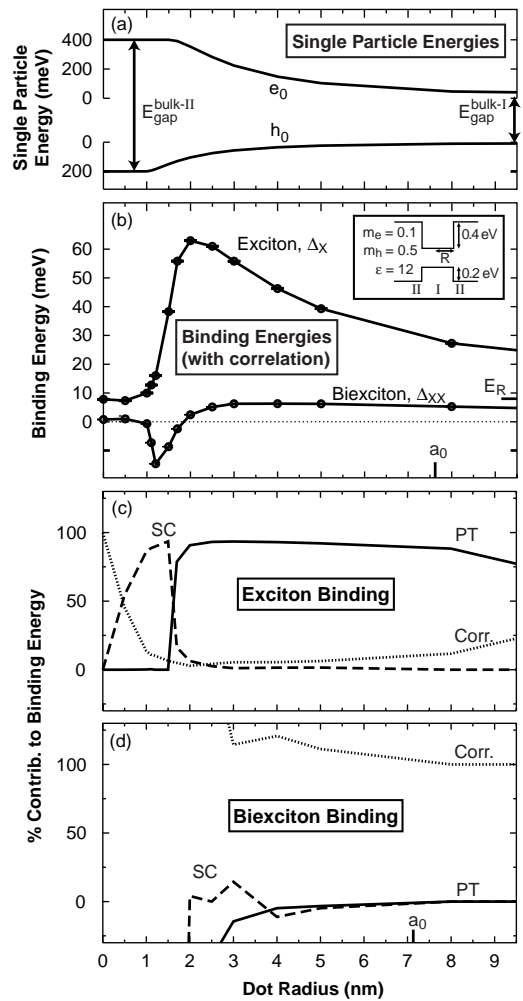}}
\caption{Exciton and biexciton binding energy versus dot radius
as calculated by QMC, for the dot geometry shown in the inset.
Panel (a) shows the energies of the non-interacting electron
and hole band edge states. Panel (b) shows the
the exciton binding energy $\Delta_X$ [Eq.~(\ref{eq:exciton})], and 
biexciton binding energy $\Delta_{XX}$ [Eq.~(\ref{eq:biex})].   The bulk exciton
Rydberg energy and Bohr radius are denoted $a_0=7.6$~nm and 
$E_R=7.9$~meV, respectively.
Contributions to exciton and biexciton binding energy 
versus dot radius are shown in (c) and (d), respectively.
Contributions are from: first order perturbation theory (PT), 
self-consistency correction (SC), and correlation (Corr.).}
\label{fig:radius}
\end{figure}

We first use a simplified single-band EMA
model which can be treated by both QMC and CI.  Our reference system
is a spherical dot with radius $R=40$~\AA, effective masses
$m_e=0.1$ and $m_h=0.5$, dielectric constant $\epsilon=12$,
and barriers $\Delta E_{\text{v}}=0.4$~eV 
and $\Delta E_{\text{c}}=0.2$~eV.
The energies of the optical and electronic 
properties of this dot are summarized in Table~\ref{table:quantities}.
We have then varied the radius from 0 to 80~\AA, while keeping the 
barriers fixed.  This yields a range of bound electron
and hole states.  The energies of the lowest (i.~e. band-edge)
states $e_0$ and $h_0$ as a function of
dot radius $R$ are shown in Fig.~\ref{fig:radius}(a).
When the radius $R$ of the dot goes to infinity we have a 3D bulk
material called ``material I'' with $m_e=0.1$, $m_h=0.5$, and
$\epsilon=12$.
When the radius $R$ of the dot goes to zero we have a 3D bulk
material called ``material II'' with $m_e$, $m_h$, and $\epsilon$
identical to ``material I.''  The band offsets between
the two materials $\Delta E_h = 0.2$~eV for the valence band
and $\Delta E_e = 0.4$~eV for the conduction band, so that the
band-gap of ``material II'' is $\Delta E_h + \Delta E_e = 0.6$~eV
larger than the band-gap of ``material I.''  The bulk exciton in
both materials is the same, and has a radius $a_0=76.2$~\AA{}, 
a binding energy $E_R=7.873$~meV. Both bulk materials have a bound 
biexciton with the same binding energy, 
$\Delta^{\text{bulk}}_{XX}=0.716$~meV $= 0.9 E_R$, (calculated by QMC).
In some calculations we have varied the barrier energy from 
$\Delta E_{\text{v}}=0.05$~eV to $\Delta E_{\text{v}}=4$~eV and
$\Delta E_{\text{c}}=0.025$~eV to $\Delta E_{\text{c}}=2$~eV,
while keeping the radius fixed at 40~\AA.
Our model system has thus been chosen to  roughly capture some
properties of small SK or colloidal dots, as summarized in
Table~\ref{table:quantities}.

\subsection{Total energies for occupation by an exciton, biexciton, 
and two-electrons}

\begin{figure*}
\includegraphics[width=\linewidth]{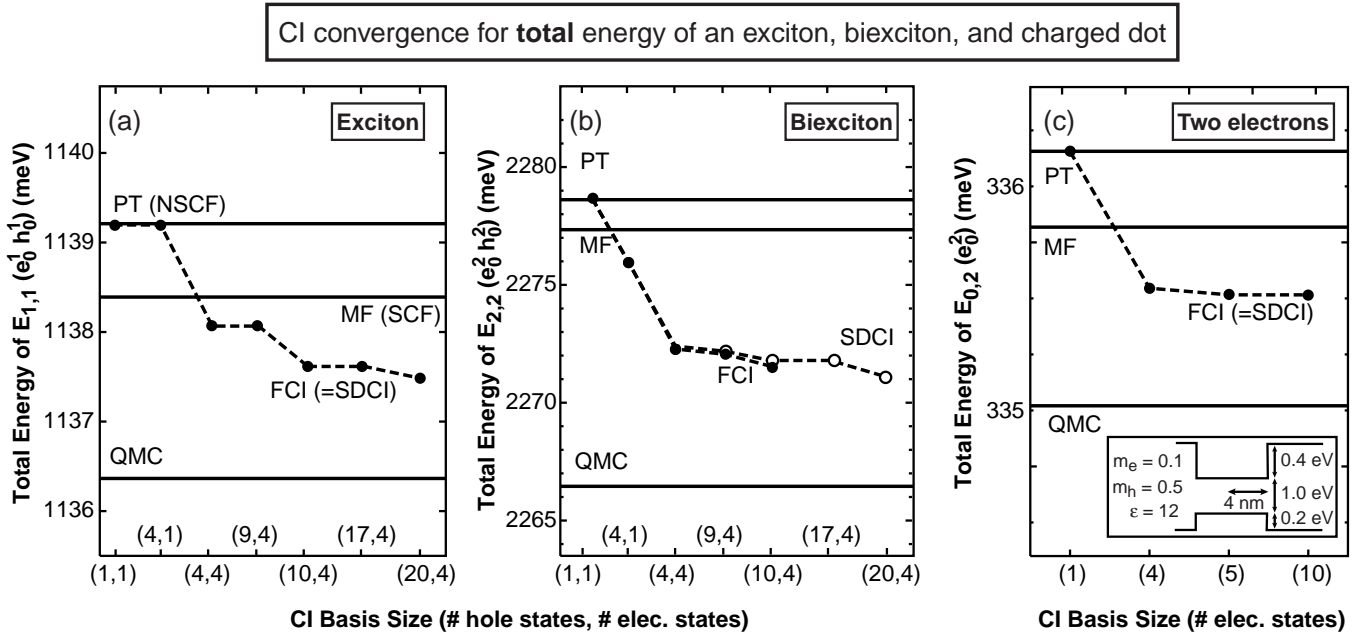}
\caption{CI convergence of the total energy for three cases: 
(a) an exciton, (b) a biexciton, and (c) two electrons.  All energies
are measured relative to the center of dot gap.
For our CI expansion, we have used single and double substitutions 
(SDCI) and also all possible determinants (FCI). Note that SDCI is
equivalent to FCI for cases (a) and (c). SDCI gives a good 
approximation to FCI for case (b), and involves far fewer determinants
(see Table~\ref{table:ndet}).
In all cases our CI expansion captures about half of the correlation energy. 
The correlation energy (and hence CI error) is a very small fraction
($< 1$\%) of the total energy in all three cases.
\label{fig:ci}}
\end{figure*}

\begin{table}
\caption{Number of determinants used for each of the CI calculations
shown in Figs.~\ref{fig:ci} and \ref{fig:ci_meas}, 
using only single and double substitutions
(SDCI), or all possible slater determinants (FCI).  Note that SDCI is 
equivalent to FCI for the case of an exciton or two electrons.
For FCI the number of CI determinants for $M$ holes and $N$ electrons occupying
$N_h$ hole states and $N_e$ electron states is 
$C_M^{2N_h}\cdot C_N^{2N_e}$.}
\label{table:ndet}
\begin{ruledtabular}
\begin{tabular}{lcrr}
& & \multicolumn{2}{c}{\# of determinants}\\
System & ($N_h$,$N_e$) & SDCI & FCI\\
\colrule
Exciton $(h^1e^1)$:& &\\
&(1,1)&4&4\\
&(4,1)&16&16\\
&(4,4)&64&64\\
&(9,4)&144&144\\
&(10,4)&160&160\\
&(17,4)&262&262\\
&(20,4)&320&320\\
Biexciton $(h^2e^2)$:& &\\
&(1,1)&1&1\\
&(4,1)&28&28\\
&(4,4)&199&784\\
&(9,4)&564&4284\\
&(10,4)&649&5320\\
&(17,4)&1356&15708\\
&(20,4)&1719&21840\\
Two Electrons$(h^0e^2)$:& &\\
&(0,1)&4&4\\
&(0,4)&28&28\\
&(0,5)&153&153\\
&(0,10)&190&190\\
\end{tabular}
\end{ruledtabular}
\end{table}

Figure~\ref{fig:ci} shows the total energy for 
(a) exciton, $E_{1,1}(e_0^1,h_0^1)$;
(b) biexciton, $E_{2,2}(e_0^2,h_0^2)$; and
(c) two-electrons, $E_{2,0}(e_0^2,h_0^0)$.  We have decomposed
the total energies into the three parts listed in Eq.~(\ref{eq:threeterms}):
first-order perturbation theory ($E_{\text{PT}}$), self-consistent
mean field ($E_{\text{PT}}+\Delta E_{\text{SC}}$), and the exact QMC result
($E_{\text{tot}}= E_{\text{PT}}+\Delta E_{\text{SC}}+\Delta E_{\text{corr}}$).
We then plot the results of CI calculations
as a function of the number of single-particle states
$(N_h,N_e)$ used to generate the CI basis set, taking either singles and
doubles only (SDCI) or all possible determinants (FCI).  
The CI energies for one determinant are equivalent to the MF result,
and the FCI values must reach the QMC result in the limit of an 
infinite basis.  The total number of CI determinants for $M$ holes 
and $N$ electrons occupying
$N_h$ hole states and $N_e$ electron states is 
$C_M^{2N_h}\cdot C_N^{2N_e}$, where $C_m^n = n!/[m!(n-m)!]$.
The factors of 2 are due to the spin-degeneracy of the single particle states. 
Table~\ref{table:ndet} lists the actual number of determinants for each of the
FCI and SDCI data points in Fig.~\ref{fig:ci}.  The first three lines
of Table~\ref{table:quantities} give a summary of the role of correlation
energy and CI convergence in the total energy of these three systems.

In each system, the total energy estimated by first-order perturbation
theory is above the true ground state energy (as required by 
the variational principle).
Self-consistency improves upon first-order perturbation theory,
and correlation provides additional improvement.
For excitons, the self-consistency decreases the energy by $\sim 1$ meV,
and correlation gives another $\sim 2$ meV improvement.  The total 
energy, however, is $E_{1,1}=1136$~meV.\ \ So, although
our CI only recovers about half of the correlation energy, the total
energy is only overestimated by about 0.1\%.
For the case of a biexciton, self-consistency also lowers the energy by
$\sim 2$ meV, while correlation lowers the energy by another $\sim 10$~meV.\ \ 
In calculations on a strain induced dot, Brask\'en et al\cite{Brasken:2000} 
found that SDCI captured
$\agt 90\%$ of the correlation energy for multi-excitons, based on comparison
to FCI for one to four excitons.  In our biexciton calculations, 
we also find that SDCI recovers nearly as much correlation energy as FCI, but
this represents only about about half of the total correlation energy. 
Again, though, correlation represents a small part of the total energy
of the biexciton, so CI (FCI and SDCI) only overestimate the 
total energy by $\sim 0.2$\%.
For a dot containing two electrons, corrections beyond first-order perturbation 
theory are much smaller, $\sim 1$~meV.\ \  In fact, for the system calculated
here, we find only a 0.35~meV decrease in the two-electron system with
self-consistency, and correlation decreases the total energy
by about another 0.8~meV.\ \  Our CI expansion again captures about half
this correlation energy, leading to negligibly small overestimation of the
total energy ($< 0.1$\%).

\subsection{Exciton and biexciton transition and binding energies}

Measured quantities such as the exciton and biexciton binding energies 
represent {\em differences} between total energies.
Even if the mean-field contributions dominate total energies,
the mean-field contributions to differences of total energies
may have significant contributions from correlation.
Lines 4-6 of Table~\ref{table:quantities}
summarize the role of correlation and CI convergence for
the exciton transition energy, $E_X$ [Eq.~(\ref{eq:extran})];
exciton binding energy, $\Delta_X$ [Eq.~(\ref{eq:exciton})]; and the
biexciton binding energy, $\Delta_{XX}$ [Eq. (\ref{eq:biex})].
Correlation is only a small part (2 meV) of the exciton
transition energy $E_X = 1136.3$~meV.\ \   So, even though our underconverged
CI fails to capture all the correlation energy, $E_X$ is only overestimated
by 0.1\%.  The same 2 meV of correlation energy is a much larger 
component of the exciton binding energy, $\Delta_X = 46.2$~meV, so
errors due to underconvergence of CI are more significant, and
CI underestimates $\Delta_X$ by more than 2\%.  The biexciton binding energy 
$\Delta_{XX} = 6.2$~meV is due entirely to 6.8~meV of correlation energy,
so CI underconvergence is much more serious.  Our CI calculation of
biexciton binding is only 65\% of the exact QMC result.

\begin{figure*}
\includegraphics[width=\linewidth]{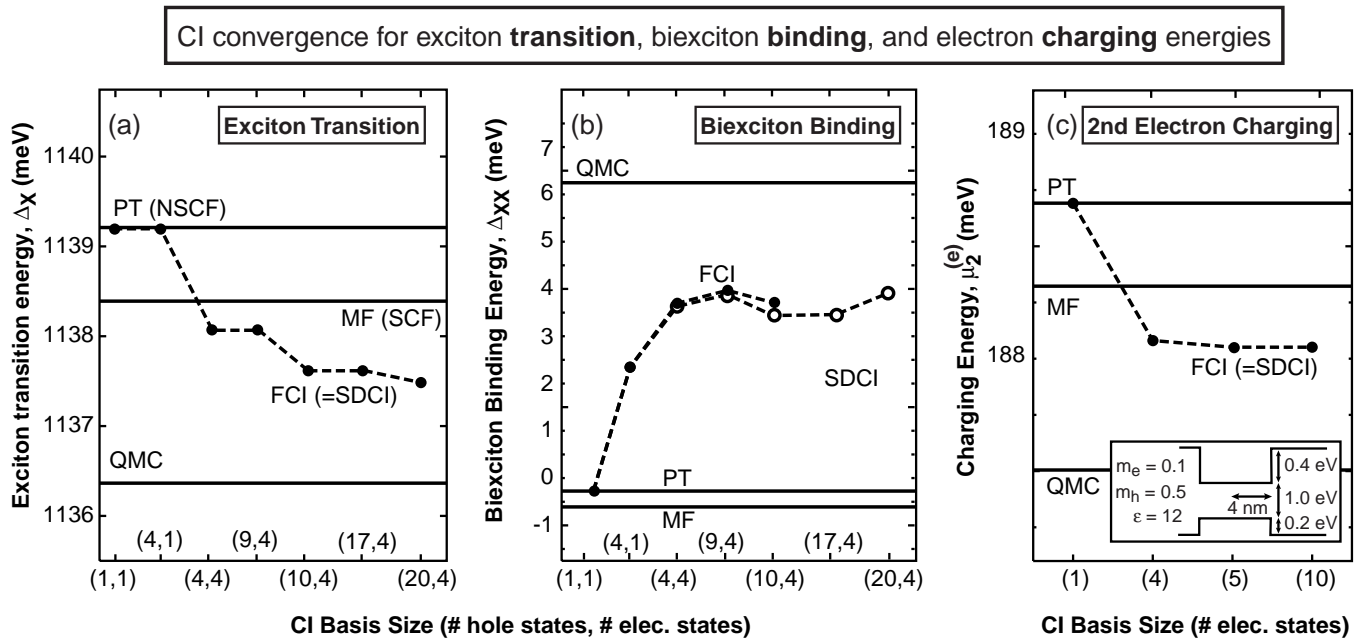}
\caption{CI convergence of:
(a) exciton transition energy, $E_X$ [Eq.~(\ref{eq:extran})];
(b)  biexciton binding energy, $\Delta_{XX}$ [Eq. (\ref{eq:biex})]; and 
(c) second electron addition energy, $\mu^{(e)}_2$ [Eq. (\ref{eq:qcharge})].
For our CI expansion, we have used single and double substitutions 
(SDCI) and also all possible determinants (FCI). Note that SDCI is
equivalent to FCI for cases (a) and (c). SDCI gives a good 
approximation to FCI for case (b), and involves far fewer determinants
(see Table~\ref{table:ndet}).
In all cases our CI expansion captures about half of the correlation energy. 
The correlation energy (and hence CI error) is a significant
fraction of the the total energy only for case (b), biexciton binding.
\label{fig:ci_meas}}
\end{figure*}

In Fig.~\ref{fig:ci_meas} we show the results of 
first-order perturbation theory ($E_{\text{PT}}$), self-consistent
mean field ($E_{\text{PT}}+\Delta E_{\text{SC}}$), the exact QMC result
($E_{\text{tot}}= E_{\text{PT}}+\Delta E_{\text{SC}}+\Delta E_{\text{corr}}$),
and CI convergence vs. basis size for
(a) the exciton transition energy and (b) the biexciton binding energy.
For the exciton transition energy, Fig.~\ref{fig:ci_meas}(a),
increasing the CI basis does improve the calculated energy,
but it is only a difference of $\sim 2$~meV out of a much larger 
exciton transition energy of 1.136~eV. 
On the other hand, the CI correction is essential to even approximate
the biexciton binding energy, shown in Fig.~\ref{fig:ci_meas}(b).  Note that
the improvement of the biexciton binding with CI basis size is not monotonic.
This is because the biexciton binding is a difference of one- and two-exciton 
energies.  As the basis is increased,  the relative improvement in
the one- and two-exciton total energies varies, thus the calculated biexciton 
binding energy can actually {\em decrease} when the CI basis is improved.
We also show the results of SDCI in Fig.~\ref{fig:ci_meas}(b).

\subsubsection{Dependence on dot size}

We have varied the dot radius from $R=0$ to $R=80$~\AA, all in the
strongly confined regime, $R\alt a_0=76.2$~\AA.
Figure~\ref{fig:radius}(b) shows the exciton and
biexciton binding energies as calculated by QMC.
Figures~\ref{fig:radius}(c) and \ref{fig:radius}(d) 
decompose the contributions to the exciton and biexciton binding
into (1) first order perturbation theory, (2) self-consistency
corrections, and (3) correlation corrections, as in Eq.~(\ref{eq:threeterms}).

The small $R$ limit is the energy of a bulk-II material, and
all excitonic binding energy is from correlation.  As the radius of the
dot increases, the bulk-II exciton binds to the dot, the 
exciton binding energy is enhanced, and most of the binding energy
comes from perturbation theory.  The maximum in the binding energy
occurs when the electron and hole are both individually bound to the
dot, but the radius is small, so that the direct Coulomb interaction
(from first order perturbation theory) is the strongest.
The exciton binding energy exhibits a clear peak at around
$R\approx 40$~\AA, in similarity with previous calculations 
by Austin.\cite{Austin:1988}   As the
dot becomes larger, the direct Coulomb interaction from perturbation
theory decreases, causing a decrease in the exciton binding energy.
Finally, as the dot becomes comparable in size to the bulk-I exciton
radius, correlation begins to have significant contributions to 
exciton binding.  In the limit $R\gg a_0$ (not shown), the binding
energy becomes that of a bulk-I exciton.

The biexciton binding energy is greatly enhanced in a quantum dot, except
for the case of a very small dot with only a single
weakly bound exciton.  We find that the biexciton binding energy is
remarkably insensitive to dot radius, having a value
$\Delta_{XX}$ between 5.1~meV and 6.2~meV ($0.7~E_R$ to $0.9~E_R$)
for dots with radii R between 2~nm and 8~nm ($0.3~a_0$ and $1.1~a_0$).
This is in contrast the exciton binding energy, $\Delta_X$, which exhibits
a clear peak at small dot radius.
The size range $10 \alt R \alt 18$~\AA{} has a negative biexciton binding.
Physically, these are small dots that can weakly bind two excitons,
but with a higher total energy than separating the two excitons on two 
non-interacting, identical dots.
We see from Fig.~\ref{fig:radius}(d) that the biexciton binding 
energy is almost entirely due to correlation, as noted before. 

\subsubsection{Dependence on barrier height}

\begin{figure}
\centerline{\includegraphics[width=0.79\linewidth]{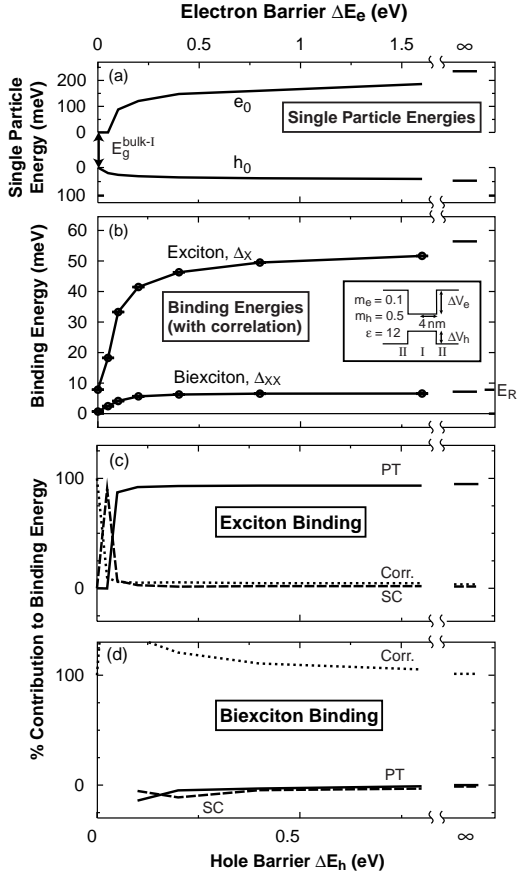}}
\caption{Exciton and biexciton binding energy (including correlation)
as calculated by QMC as a function of 
versus barrier energy, with the constraint $\Delta E_e/\Delta E_h =2$,
for the dot geometry shown in the inset.  Panel (a) shows the single-particle
energies of the non-interacting electron and hole band edge states.   
Panel (b) shows the the exciton binding energy $\Delta_X$ 
[Eq.~(\ref{eq:exciton})] and biexciton binding energy $\Delta_{XX}$ 
[Eq.~(\ref{eq:biex})].   The bulk exciton Rydberg energy is 
denoted $E_R = 7.9$~meV.
Contributions to exciton and biexciton binding energy 
versus barrier energy are shown (c) and (d), respectively.
Contributions are from: first order perturbation theory (PT), 
self-consistency correction (SC), and correlation (Corr.).}
\label{fig:barrier}
\end{figure}

To study the effect of finite confining barriers on exciton and
biexciton binding energies, we have varied the dot barriers from
zero to infinity.  In all calculations we have kept $\Delta E_e
/\Delta E_h = 2$ and used a radius of 40~\AA.
In Fig.~\ref{fig:barrier}(b) we plot the binding energies of excitons 
and biexcitons calculated with QMC as a function of barrier height.
The 40~\AA{} dot is able to bind an electron once $\Delta E_e \agt 30$~meV,
and binds a hole once $\Delta E_h \agt 5$~meV.
Unlike the behavior seen with varying the dot radius, increasing
the confining potential leads to a monotonic increase in exciton
and biexciton binding energies.  For zero barrier potential, the exciton 
has the bulk-I exciton binding energy, $\Delta_X = E_R^{(I)} = 7.9$~meV. 
As the barrier potential is increased enough to
bind both electrons and holes, the exciton binding
increases rapidly.  The binding energy reaches 
a maximum of $\Delta_X = 55~\text{meV} =  7 E_R$  for infinite barriers.
Similarly, the biexciton binding energy starts from the bulk biexciton
binding energy $\Delta_{XX} = 0.7~\text{meV} = 0.1 E_R$ and increases
to a maximum of $\Delta_{XX} = 7.2~\text{meV} = 1.0 E_R$ for infinite
barriers.  Figures~\ref{fig:barrier}(c) and \ref{fig:barrier}(d) show 
the contributions of 
perturbation theory, self-consistency correction, and
correlation to the exciton and biexciton binding energy.
Except for very weakly confined dots, the exciton is very well described
by first-order perturbation theory.  For weak confinement, the electron
is unable to bind, but self-consistent interaction with the hole is
able to bind the electron, so that the exciton binding energy is
almost entirely due to self-consistency.  For the weakest confinement, neither
the electron nor the hole is bound, and the excitonic binding is entirely
due to correlation.  Again, biexciton binding is due entirely
to correlation.

\subsection{Multi-exciton energies}

\begin{figure}
\includegraphics[width=\linewidth]{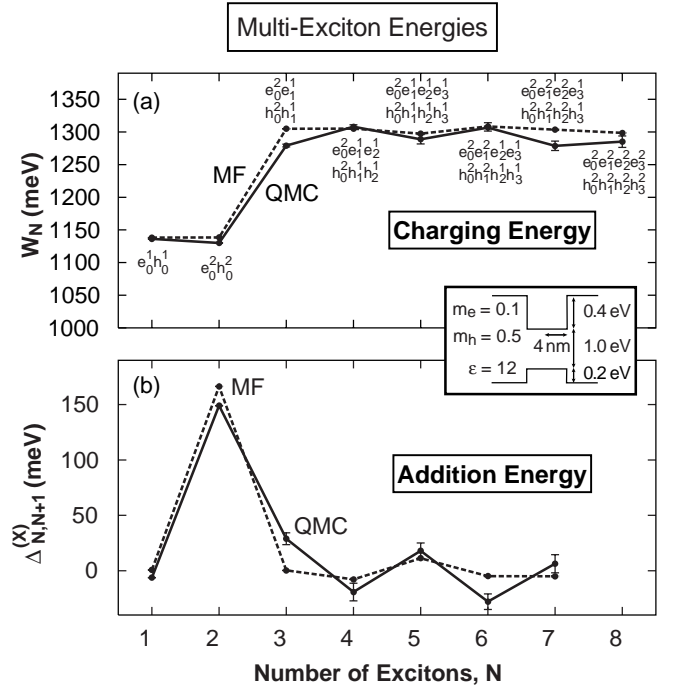}
\caption{(a) Exciton charging [Eq.~(\ref{eq:ncharge})], 
and (b) addition energies
[Eq.~(\ref{eq:nadd})], for the dot shown in the inset.
Because excitons are neutral, it is energetically favorable for 
a dot to hold many excitons.
\label{fig:exadd}}
\end{figure}

Figure~\ref{fig:exadd} shows mean-field and exact (QMC) results for
the multi-exciton charging energies $W_N$
[Eq.~(\ref{eq:ncharge})], and the multi-exciton additions energies, 
$\Delta^{(X)}_{N,N+1}$ [Eq.~(\ref{eq:nadd})]. 
The most prominent feature is the jump in the charging energy
for $W_3$, which also appears as a peak in the addition energy
$\Delta^{(X)}_{2,3}$.  This ``shell effect'' arises because
only the first two excitons can occupy the lowest energy
$e_0$ and $h_0$ states.  Starting with the third exciton, Pauli
exclusion requires the addition excitons to start filling
the next energy shell, $e_1h_1$ through $e_3h_3$.  This is
a feature of the single particle model, and does not require
any treatment of correlation.
Correlation is necessary to describe the decrease in
charging energy for the second exciton, $W_2<W_1$, or equivalently
the negative value of the first exciton addition energy 
$\Delta^{(X)}_{1,2}=-6.2$~meV.  This is the
positive biexciton binding energy $\Delta_{XX} = 6.2$~meV,
discussed earlier.  As shown in
lines 7-9 of Table~\ref{table:quantities}, the correlation contribution
for the second charging energy $W_2$ is 8.9~meV, considerably larger
than the 2.0~meV for $W_1$.  Our CI only captures about half
the correlation energy, so it slightly overestimates the exciton
charing energies, and considerably underestimates the negative
value of $\Delta^{(X)}_{1,2}$.

\subsection{Electron loading energies}

\begin{figure}
\includegraphics[width=\linewidth]{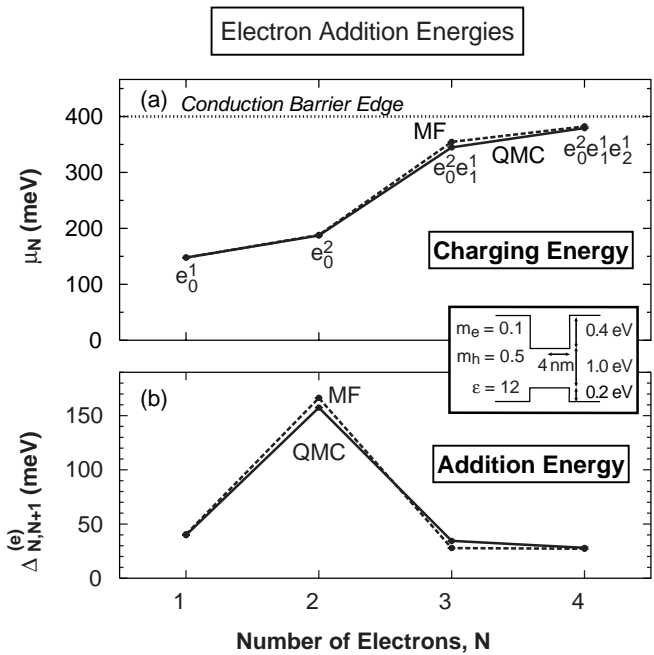}
\caption{(a) Electron charging energies  [Eq.~(\ref{eq:qcharge})],
and (b) addition energies [Eq.~(\ref{eq:qadd})], for the dot shown in the inset.
This dot can only hold up to four electrons, due to Coulomb repulsion.
The conduction band minimum energy of the barrier material, 
$\Delta E_e=400$~meV is shown in panel (a).
\label{fig:qadd}}
\end{figure}

Figure~\ref{fig:qadd} shows mean-field and exact (QMC) results for
electron charging energies $\mu_N$,
[Eq.~(\ref{eq:qcharge})], and the electron additions energies, 
$\Delta^{(e)}_{N,N+1}$ [Eq.~(\ref{eq:qadd})]. 
Because electrons are charged, Coulomb
repulsion quickly limits the number of electrons that can be loaded
into the dot.  For our model, shown in the inset to Fig.~\ref{fig:qadd}, it is
only energetically favorable to add four electrons; beyond this, electrons
would rather escape into the barrier material conduction band,
shown as a dashed horizontal line in Fig.~\ref{fig:qadd}(a).
There is a peak in the electron addition energy $\Delta^e_{2,3}$ in
Fig.~\ref{fig:qadd}(b).  This is due the filling of the $e_0$ state
by a spin-up and spin down electron (another ``shell effect'').
Both QMC and MF capture this single particle effect.
As shown in Fig~\ref{fig:ci_meas}(c),
our CI expansion recovers about half of the correlation energy for
two electrons.  However, the correlation energy in a two-electron
dot is only about 1~meV, so CI errors are a negligible 0.5~meV.
The small value of correlation and the good agreement of our CI
calculations for dot charging are summarized in 
last three lines of Table~\ref{table:quantities}

\section{Application of CI to a multi-band dot described via plane-wave
pseudopotentials}
\label{sec:CIEPM}

QMC calculations are currently limited to either small systems containing
up to a few hundreds of electrons,\cite{Mitas:2000,Kent:1999,Torelli:2000}
or to highly simplified model Hamiltonians (such as the EMA).
A more accurate description of the electronic structure (Fig.~\ref{fig:methods})
of semiconductor quantum dots can be obtained using the pseudopotential
approach.\cite{Wang:2000a}  Unfortunately, QMC methods are presently unable to
deal with the large number of electrons of a typical quantum dot,
and CI is the only viable approach to
treat correlation effects in large quantum dots described
by atomistic pseudopotentials. 
In addition, the diagonalization of the CI Hamiltonian
gives access to the excited states (unavailable 
in ground-state QMC calculations) as well as the ground state
of the electronic system, thus enabling the calculation of
the optical spectrum of quantum dots.

In order to illustrate the capabilities of the CI approach
combined with a pseudopotential description of the electronic structure,
we consider a nearly spherical CdSe quantum dot having the wurtzite lattice
structure and a diameter of 38.5~\AA.
The surface dangling bonds are fully passivated using ligand-like 
atoms.\cite{Wang:1998a}
This quantum dot is representative of CdSe nanocrystals
grown by colloidal chemistry methods.

We consider here only low-energy excitations of the
electronic system, which are obtained by promoting
electrons from states near the top of the valence band to 
states near the bottom of the conduction band. 
The band-edge solutions of Eq.~(\ref{eq:se}) can be efficiently
obtained using the folded spectrum 
method,\cite {Wang:1994a,Wang:1994c,Wang:1996a}
which allows one to calculate {\it selected} eigenstates
of the Schr\"odinger equation with a computational
cost that scales only linearly with the size of the system.
In this approach, Eq. (\ref{eq:se}) is replaced by the folded-spectrum equation

\begin{widetext}
\begin {equation} \label {Fse}
\left [ - \nabla ^2 + 
V_{ps} ({\bf r}) + {\hat V} _{NL} 
- \varepsilon_{ref} \right ]^2 \psi_i ({\bf r}, \sigma) = 
(\varepsilon^0_i - \varepsilon_{ref})^2 \; \psi_i ({\bf r}, \sigma) \, , 
\end {equation}
\end{widetext}

\noindent where $\varepsilon_{ref}$ is an {\it arbitrary}
reference energy.  The lowest energy eigenstate of Eq. (\ref{Fse}) 
coincides with the solution of 
the Schr\"odinger equation [Eq. (\ref{eq:se})]
whose energy is closest to the reference energy $\varepsilon_{ref}$.
Therefore, by choosing the reference energy in the band gap, 
the band edge states can be obtained by minimizing the functional 
$A[\psi] = \langle \psi | ({\hat H} - \varepsilon_{ref})^2 | \psi \rangle $.

The solution of Eq. (\ref {Fse}) is performed by expanding
the wave functions $\psi_i ({\bf r}, \sigma)$ in a plane-wave basis set.
To this purpose, the total pseudopotential $V_{ps} ({\bf r})$ is defined
in a periodically repeated supercell $\Omega$ containing the 
quantum dot and a portion of the surrounding material.
The supercell $\Omega$ is sufficiently large to ensure that the solutions
of Eq. (\ref {Fse}) are converged within 1 meV.
The single-particle wave functions can then be expanded as 
$\psi_i ({\bf r}, \sigma) = \sum_{\bf G} \, c_i ({\bf G}, \sigma) \, 
\exp (i{\bf G} \cdot {\bf r}) $,
where the sum runs over the reciprocal lattice vectors ${\bf G}$
of the supercell $\Omega$. The energy cutoff of the plane-wave
expansion is the same used to fit the bulk electronic
structure, to ensure that the band-structure consistently approaches
the bulk limit.
The minimization of the functional $A[\psi]$ is carried
out in the plane-wave basis set using a 
preconditioned conjugate-gradients algorithm.

In the next step we construct a set of Slater
determinants $|\Phi_{h_1 \cdots h_N , e_1 \cdots e_N} \rangle$
[see Eq. (\ref{eq:configuration})] obtained by creating $N$ holes in the
valence band and $N$ electrons in the conduction band,
and diagonalize the CI Hamiltonian in this basis set.
Using the CI approach, 
we have calculated the multiexciton spectrum  of a CdSe dot. 
We consider here up to three excitons and we use a CI basis
set of 480 configurations for the single exciton,
43890 configurations for the biexciton,
and 20384 configurations for the tri-exciton. 
All the relevant interactions (including electron-hole exchange)
are included in the CI calculations.
We assume that when an $N$-exciton is created in the quantum
dot, it relaxes non-radiatively to the ground state before decaying radiatively
into an $(N-1)$-exciton. 

\begin{figure}
\includegraphics[width=\linewidth]{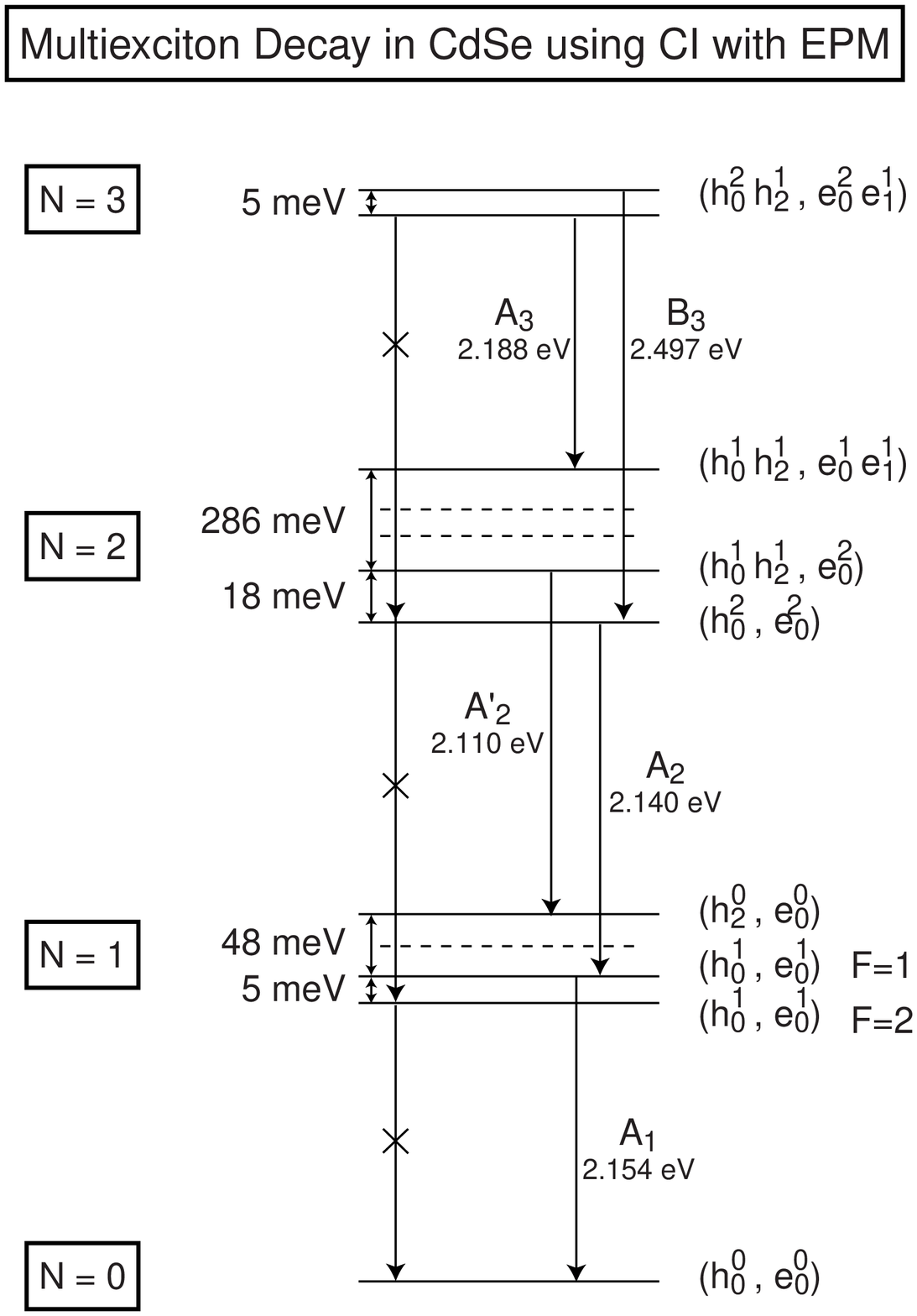}
\caption{Schematic illustration of the leading contributions to
peaks ($A_3, B_3, A'_2, A_2, A_1$) appearing in Fig.~\ref{fig:cdse}.
Solid horizontal lines are energies of $N=0$ to $N=4$ excitons, with
dashed lines indicating states that do not participate in dipole transitions.}
\label{fig:cdse2}
\end{figure}

\begin{figure}[b]
\centerline{\includegraphics[width=0.85\linewidth]{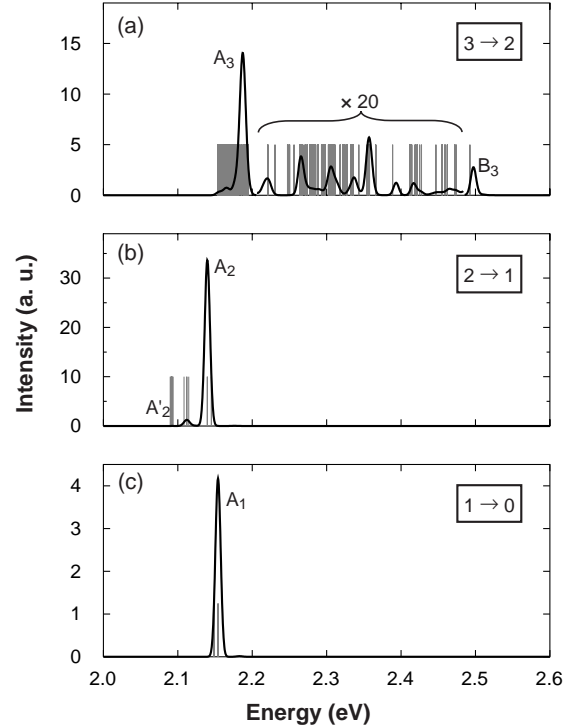}}
\caption{Exciton transition energies for a CdSe dot for (a)
decay from three to two excitons, (b) decay from two to one excitons,
and (c) decay of a single exciton.  The intensity scale is different
in each of the three panels, and weak transitions between peaks
$A_3$ and $B_3$ in (a) have been magnified by $\times 20$.
Grey vertical lines indicate all calculated transition energies,
and solid black lines in the Gaussian broadened transitions weighted by
calculated dipole transition strengths.}
\label{fig:cdse}
\end{figure}

The calculated multiplet levels are shown in Fig.~\ref{fig:cdse2}
and the emission spectrum is shown in 
Fig.~\ref{fig:cdse}. The three panels of Fig.~\ref{fig:cdse} correspond to
the recombination of (a) a tri-exciton into a biexciton ($3 \rightarrow 2$),
(b) a biexciton into a single-exciton ($2 \rightarrow 1$), 
and (c) a single-exciton into the ground state ($1 \rightarrow 0$), 
respectively.  We assume that the low-energy states
of the $N$-exciton are thermally populated ($kT = 5 \; {\rm meV}$)
before recombination.
We see from Fig.~\ref{fig:cdse} that:

(i) The single-exciton recombination spectrum, Fig.~\ref{fig:cdse}(a),
shows a single peak ($A_1$) centered at $2.154 \; {\rm eV}$. 
It is well known \cite{Nirmal:1995} that in CdSe nanocrystals
the electron-hole exchange interaction splits the lowest-energy 
excitonic state ($h_0^1 , e_0^1$) into two doublets, having total 
angular momentum $F=2$ and $F=1$ respectively (see Fig.~\ref{fig:cdse2} ).
The lower-energy doublet ($F=2$) is optically forbidden,
while the higher-energy doublet ($F=1$) is optically allowed.
We find an energy separation of $\sim 5 \; {\rm meV}$ between
the two doublets. The emission peak $A_1$ observed in Fig.~\ref{fig:cdse}
comes from the recombination of the higher-energy doublet,
which is thermally populated. This explains
the relatively weak intensity of the single-exciton peak.

(ii) The biexciton recombination spectrum, Fig.~\ref{fig:cdse}(b), 
shows a strong peak ($A_2$) centered at $2.140 \; {\rm eV}$.
This peak originates from the recombination of a biexciton in the ground state
($h_0^2, e_0^2$) into a single exciton in the $F=1$ state.
The weak shoulder to the red of the main peak ($A'_2$) is due to
the recombination of a thermally occupied higher-energy
biexciton state in the configuration ($h_0^1 h_2^1 , e_0^2$).
Note that several transitions from the biexciton ground state
to single-exciton excited states are in principle possible,
but have very weak oscillator strength. These transitions
would occur to the {\it red} of the fundamental transition.
The calculated biexciton binding energy 
is $ 2E_X - E_{XX} \sim 4 \; {\rm meV}$.
This value is probably underestimated due to the under-convergence
of the CI expansion. 
Interestingly, the ``apparent" biexciton binding energy,
i.e. the red-shift of the main biexciton peak $A_2$ with
respect to the single-exciton peak $A_1$, is $ \sim 14 \; {\rm meV}$
(not $4 \; {\rm meV}$).
The reason is that the biexciton recombination
takes the quantum dot in the $F=1$ excited state,
rather than the $F=2$ ground state (see Fig.~\ref{fig:cdse2}). 
Thus we have 
$E(A_1) - E(A_2) = (E_X^{F=1} - E_{0,0}) - (E_{XX} - E_X^{F=1}) =  
\Delta_{XX} + 2 \, (E_X^{F=1} - E_X^{F=2}) = 
4 + 2 \times 5 \; {\rm meV} = 14 \; {\rm meV}$.

(iii) In the case of three excitons
we find that the ground state wave function originates
primarily from the non-Aufbau configuration
$h_0^2 h_2^1 ; e_0^2 e_1^1$. In fact, the third hole
prefers to occupy the $p$-like $h_2$ state rather
than the $s$-like $h_1$ state, due to reduced Coulomb repulsion
with the remaining two holes.
Two main transitions are possible from the three-exciton ground
state: the $e_0 \rightarrow h_0$ recombination, which leaves
the system into the excited biexciton configuration $h_0^1 h_2^1 ; e_0^1 e_1^1$,
leads to peak $A_3$ located at $2.188 \; {\rm eV}$. 
The $e_1 \rightarrow h_2$ recombination,
which takes the system into the ground-state biexciton
configuration $h_0^2 ; e_0^2$, is responsible for peak $B_3$ 
centered at $2.497 \; {\rm eV}$.
Note that the $B_3$ transition originates from an exchange-split
tri-exciton state (see Fig.~\ref{fig:cdse2}) which is thermally populated,
hence the relatively weak oscillator strength of the $B_3$ transition.

Note that a calculation considering only ground-state
to ground-state transitions would miss
most of the peaks observed in Fig.~\ref{fig:cdse}. The capability
of the CI expansion to access excited states,
coupled with the possibility of using a multi-band
pseudopotential Hamiltonian for the calculation
of the single-particle energies and wave functions,
makes it the method of choice for calculating
excited states of semiconductor quantum dots.

\section{Conclusion}
We have studied the effects of correlation on a simplified, single-band
model dot using both QMC and CI, and have studied correlation in
the multi-exciton PL spectra of a realistically modeled CdSe dot
using CI.
Our results for the simplified, single band model are summarized in
Table~\ref{table:quantities}.  We find the following results for our model:
(1) total energies for an exciton, biexciton, and two electrons
are dominated by mean field effects, so that correlation energies
and CI convergence errors are less than 1\% [see Fig.~\ref{fig:ci}];
(2) typical exciton transition energies, which
are $\sim 1$~eV, can be calculated to closer than 1\% by
perturbation theory, with only a $\sim 2$~meV correlation correction
[see Fig.~\ref{fig:ci_meas}(a)];
(3) typical exciton binding energies are $\sim 46$~meV, with only 2~meV
from correlation, and our CI captures roughly half of the correlation
to give exciton binding energies that are nearly 98\% of the exact QMC value;
(4) typical biexciton binding energies are positive $\sim 6$~meV,
almost entirely due to correlation energy,  and our CI only recovers
about 65\% of the exact QMC value [see Fig.~\ref{fig:ci_meas}(b)];
(5) exciton charging energies are $\sim 1130$~meV 
and well described by CI, while exciton addition energies can be due
entirely to correlation, in which case our CI is only qualitatively
correct; and (6) typical electron
charging energies are $\sim 150$~meV, of which correlation contributes
very little ($\sim 1$~meV), likewise, electron addition energies
are $\sim 40$~meV with very little correlation contribution, so
that CI is accurate to about 1-2\% for electron addition energies.

Although QMC is a good method for testing convergence of CI
on a simplified, single band model, only CI may be used  on our
more realistic model of CdSe.  Our multi-band pseudopotential
model capture the correct symmetries and electronic structure of the
dots, leading to qualitatively different predictions than single-band
models.  For example, the multiplet structure presented in 
Fig.~\ref{fig:cdse2} requires a multi-band description of the single
particle levels.  Some of the details of our realistic CdSe calculation that
are missing from our single-band CI model are:
(1) different degeneracies of the single-particle hole levels due to a 
multi-band description of the valence band states,
(2) electron-hole exchange splitting of 5 meV in the ground
state $(h_0^1,e_0^1)$ exciton, (4) the existence of many
weak transitions that are symmetry forbidden in single band models,
An additional benefit of CI is that it gives excited state energies 
necessary to identify some of the peaks that appear in single-dot
photoluminescence spectra.

We conclude that correlation effects are important to some quantities,
such as exciton binding and exciton addition energies, and essential
to calculate positive binding energies.  QMC methods are well-suited
for simple, single-band models.  Applications to realistic
models which capture the proper symmetries and electronic structure
of quantum dots are currently restricted to CI methods.  We find
that CI calculations including all bound states are accurate to better
than 3\% for many measurable properties, as listed in 
Table~\ref{table:quantities}.  Even for biexciton binding, which is
dominated by correlation, our CI calculations are qualitatively correct,
capturing about 65\% of the QMC prediction for a simplified models.  Therefore
we conclude that realistic multi-band models combined with
perturbation theory and a judicious use of CI for correlation corrections
is a computational approach well-suited to realistic modeling of interacting
electrons and holes in SK and colloidal semiconductor quantum dots.

\acknowledgments
Work supported by the DOE Office of
Science --  Basic Energy Sciences, Division of Materials Sciences under
contract No. DE-AC36-99GO10337.

\bibliography{dots}
\end{document}